\documentclass[pra,reprint,a4paper,twocolumn,nofootinbib,showpacs,superscriptaddress,showkeys]{revtex4-1}
  \usepackage{amsmath,amssymb,amsfonts}
  \usepackage[dvips]{graphicx}
  \usepackage{graphicx}
  \usepackage{txfonts}
  \usepackage{textcomp}
  \usepackage{graphicx}
  \usepackage{epsfig}
  \usepackage{upgreek}
  \usepackage{braket}
  \usepackage{bbold}
  \usepackage{psfrag}
  \usepackage[mathscr]{euscript}
  \usepackage[unicode]{hyperref}
  \hypersetup{
      unicode=true,                                           
      a4paper=true,
      plainpages=false,
      pdffitwindow=true,                                      
      pdftitle={Soliton Generation},                      
      pdfauthor={Simon Alexander Gardiner},                            
      pdfsubject={},                                          
      pdfkeywords={Bose--Einstein, condensate, BEC, solitons},
      colorlinks=true,                                        
      linkcolor=blue,                                         
      citecolor=blue,                                         
      filecolor=blue,                                         
      urlcolor=blue                                           
  }
\begin{document}

\title{Noise-Free Generation of Bright Matter-Wave Solitons}
\author{M. J. Edmonds}
\affiliation{Quantum Systems Unit, Okinawa Institute of Science and Technology Graduate University, Onna, Okinawa 904-0495, Japan}
\author{T. P. Billam}
\affiliation{Joint Quantum Center (JQC) Durham-Newcastle, School of Mathematics, Statistics and Physics,
Newcastle University, Newcastle upon Tyne NE1 7RU, United Kingdom}
\author{S. A. Gardiner}
\affiliation{Joint Quantum Center (JQC) Durham-Newcastle, Department of Physics, Durham University, Durham DH1 3LE, United Kingdom}
\author{Th.\ Busch}
\affiliation{Quantum Systems Unit, Okinawa Institute of Science and Technology Graduate University, Onna, Okinawa 904-0495, Japan}
\date{\today}

\begin{abstract}
We show how access to sufficiently flexible trapping potentials could be exploited in the generation of three-dimensional atomic bright matter-wave solitons. Our proposal provides a route towards producing bright solitonic states with good fidelity, in contrast to, for example, a non-adiabatic sweeping of an applied magnetic field through a Feshbach resonance.
\end{abstract}
  
\maketitle

\section{Introduction}
\noindent The toolbox of ultracold atomic physics provides opportunities to understand phenomena intrinsic to these systems --- as well as the unprecedented ability to investigate effects associated with other physical models through ideas from quantum simulators \cite{georgescu_2014}. Ultracold atomic gases furthermore offer the flexibility to access novel physical effects and parameter regimes, due to their high experimental controllability. In particular, it is now feasible to engineer the dimensionality \cite{garraway_2016}, the strength of the particle interactions \cite{chin_2010} and the potential landscape \cite{dunlop_2017,boyer_2006} of these systems to an almost arbitrary degree.

Under typical experimental conditions trapped ultracold atomic gases with weak repulsive interactions are unconditionally stable, but in the case of attractive interactions, where this is not the case, bright solitary waves have also been generated. Early experimental work focused on generating single \cite{khaykovich_2002,cornish_2006} as well as trains \cite{strecker_2002} of bright solitons and more recently interest has focussed on using bright solitons as an experimental probe for potential barriers \cite{marchant_2013,marchant_2016}. Interpreting the observed stability of these states in terms of their relative phase \cite{nguyen_2014,nguyen_2017}, and also their scattering dynamics in the presence of disorder \cite{lepoutre_2016,boisse_2017} have also been topics of recent interest.

In the context of ultracold atomic gases, bright matter-wave solitons have been shown to possess a number of unique features. In particular, regions of chaotic dynamics have been identified \cite{martin_2007,martin_2008} for trapped solitons. Bright solitons' particle-like nature has yielded particle models for the center-of-mass dynamics of these systems in harmonic \cite{khawaja_2002} as well as periodic \cite{poletti_2008} potentials. Complementary to this, bright solitary matter waves have been suggested as strong candidates for atomic interferometry \cite{martin_2012,helm_2015,helm_2018} which has led very recently to the first realization of a bright soliton based matter-wave interferometer with a cloud of $^{85}$Rb atoms \cite{mcdonald_2017}. It  has also been suggested that bright solitons could be used for the generation of Bell states for quantum information processing \cite{gertjerenken_2013}, as well as for quantum thermodynamics \cite{li_2018} applications. Proposals for controllably splitting matter-wave condensates to generate bright solitons with fixed relative phase also exist \cite{billam_2011}, and  the generation of so-called ``breathers,'' excited states of an attractively interacting gas, has also been realized experimentally \cite{everitt_2015}. Attractive interactions can also facilitate the formation of molecule-like states comprised of several solitons \cite{khawaja_2011}, while understanding the behaviour of bright soliton states in the non-integrable context has led to the identification of novel dynamics \cite{dingwall_2018}.

The experimental generation of bright matter wave solitons generally relies on being able to tune the scattering length of the condensate using Feshbach resonances. The condensate is typically created on the repulsive side of the resonance, and a sudden switch into the attractive regime is then used to create the solitons. However, this non-adiabatic process can lead to significant heating and losses of the atomic ensemble, and therefore allows only limited control over the final size and state of the soliton.

\begin{figure}[b]
\includegraphics[scale=0.5]{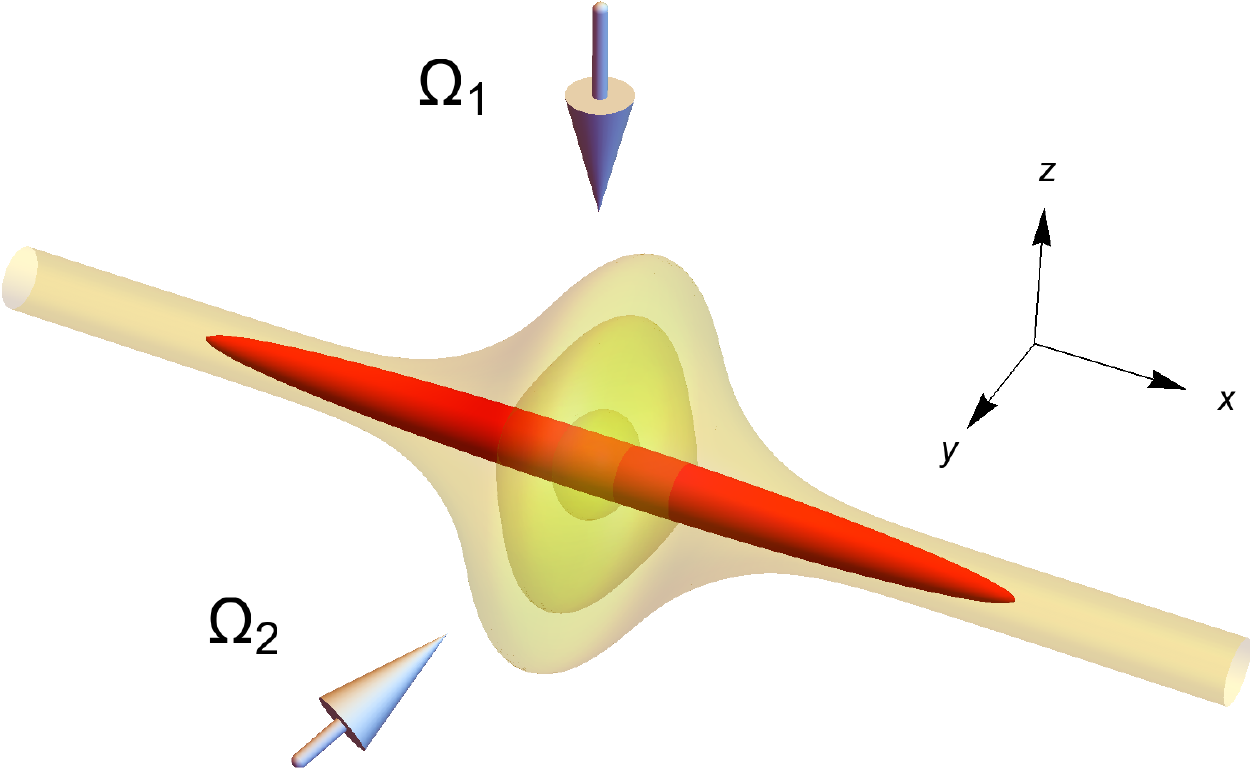}
\caption{\label{fig:Schematic}(Color online) Schematic representation of the soliton engineering protocol. The yellow surfaces represent contours of constant energy of the tailored trapping potential [see Eq.~\eqref{eqn:3dpot} and \eqref{eqn:sol3dpot}]. The red ellipsoid is the resulting ground state soliton. The two arrows are indicative of two lasers $\Omega_{1}$, $\Omega_{2}$ used to form for example a hypothetical three-dimensional painted trapping potential.}
\end{figure} 

In this work we suggest one possible solution for this problem, and how it should be possible to generate bright solitonic states with good fidelity. Our proposal is related to the ideas discussed in the area of ``shortcuts to adiabaticity,'' and in particular the so-called fast-forward theory, where dynamical processes are designed in such a way to ensure an adiabatic evolution in finite time \cite{masuda_2009,masuda_2008}. In fact, we suggest to choose the external potentials in such a way that the stationary condensate mode before and after the Feshbach switch, in conjunction with an appropriate switch of the external potential, is the same. On the attractive side the interaction and the kinetic energy terms act in opposition and compensate each other (in principle no external potential is necessary), however on the repulsive side the kinetic and the interaction terms act together, which means an external potential is required to localise the condensate. Our central idea is to design this external potential in such a way that the relevant stationary states are the same in both cases.

To engineer the target state, we need to construct a carefully tailored trapping potential. In recent years there has been enormous progress in the construction of arbitrary trapping potentials, for example via time-averaged or ``painted'' optical dipole potentials \cite{henderson_2009}. Very recently holographic potentials \cite{fujita_1996,morinaga_1996,bruce_2011} generated by spatial light modulators \cite{nogrette_2014} have been realized, as well as recent experimental work realizing three-dimensional optical tweezers for manipulating ensembles of cold atoms \cite{chisholm_2018}. While such arbitrary trapping potentials have typically been considered in two dimensions, it does not seem unreasonable to expect that comparable control over three dimensional trapping potentials is within reach. Figure \ref{fig:Schematic} shows the soliton engineering protocol schematically. Here, the tailored initial trapping potential, [see Eqs.~\eqref{eqn:3dpot} and \eqref{eqn:sol3dpot}] is represented by yellow isosurfaces of constant energy. The resulting ground state soliton is pictured in red. 

The paper is organized as follows. In Section \ref{sec:pe} we introduce our protocol using a simple pedagogical example to draw out the core features of our scheme. Following this, in Section \ref{sec:se} we explore the robustness of the scheme in three dimensions using numerical simulations, and interrogate various aspects of the fidelity of states generated using our method. In Section \ref{sec:q1d} we investigate the effect of performing an interaction quench on a trapped bright soliton. in Section \ref{sec:fluc} we study the role of fluctuations using the Truncated Wigner methodology. We then proceed to look at finite-time interaction changes in Section \ref{sec:qd}. We summarize our finding in the conclusion, Section \ref{sec:con}.     

\section{\label{sec:pe}Pedagogical example}
\noindent To clearly illustrate our process, we first consider an effective one-dimensional pedagogical model, where the radial dynamics of an atomic cloud are assumed to be frozen out. In this case the condensate is described by a time-independent, one-dimensional Gross--Pitaevskii equation of the form
\begin{equation}\label{eq:1DGPE}
\mu\psi=-\frac{\hbar^{2}}{2m}\frac{d^{2}\psi}{d x^{2}} + U\psi + g_{\mathrm{1d}}N|\psi|^{2}\psi,
\end{equation}
where $\mu$ defines the chemical potential, $m$ is the atomic mass, $N$ is the number of particles and $g_{\mathrm{1d}}=2\hbar\omega_{\rho}a_{s}$ accounts for the interaction strength between the particles, with $\omega_\rho$ the transverse harmonic trapping frequency and $a_s$ the $s$-wave scattering length. In writing the quasi-one-dimensional interaction parameter $g_{\mathrm{1d}}$, we have dimensionally reduced the three-dimensional Gross--Pitaevskii equation by assuming tight radial trapping such that the radial part of the wave function is $\psi_{R}(\rho)=(1/a_\rho\sqrt{\pi})\exp(-\rho^2/2a_{\rho}^{2})$. The external potential is described by $U(x)$ and we assume that different potential shapes can be created using optical painting or SLM techniques. We aim to have a situation where, before and after the Feshbach switch from positive to negative scattering length, at the same magnitude, the condensate wavefunction, $\psi(x)$, takes the form of a stationary  bright soliton 
\begin{equation}
\psi_{S}(x) = \frac{1}{2\sqrt{\ell_{0}}}\mathop{\mbox{sech}}(x/2\ell_{0}),
\label{Eq:SolitonWavefunction}
\end{equation}
with $\ell_{0}=\hbar^{2}/mg_{\mathrm{1d}} N$. Note that having the initial (positive) and final (negative) scattering lengths have the same magnitude is not an absolute requirement, however it is also not physically unreasonable and substantially simplifies the discussion. It will also be seen later on [see Secs.~\ref{sec:se} and \ref{sec:q1d}] that the soliton length scale can have a more complicated dependence on the interactions, however here our discussion focuses on one simple example. For the attractive regime this is automatically fulfilled if the trapping potential is switched off, $U(x)=0$, as the state given by Eq.~\eqref{Eq:SolitonWavefunction} is then an exact stationary state. However for the repulsive setting one needs to determine the external potential associated with a lowest-energy stationary state of this exact form. While it is not obvious for a general target state that such a potential even exists, in our case it can be straightforwardly calculated by realising that $\psi_S(x)$ is node-free, and then solving Eq.~\eqref{eq:1DGPE} as
\begin{align}
\begin{split}
U(x)-\mu =&\frac{\hbar^{2}}{2m}\frac{1}{\psi_{S}}\frac{d^{2}\psi_{S}}{dx^{2}}-g_{\mathrm{1d}}N|\psi_S|^{2} \\
=&\frac{1}{4\ell_{0}^{2}}\left[\frac{\hbar^{2}}{2m}-\left(\frac{\hbar^{2}}{m} + g_{\mathrm{1d}}N\right)\mathop{\mbox{sech}^{2}}(x/2\ell_{0})\right].
\end{split}
\end{align}
Here the chemical potential of the bright soliton is given by $\mu=-mg_{\mathrm{1d}}^{2}N^2/8\hbar^2$. One can immediately see that this effectively just confirms the known result that the ground state wavefunction of a single one-dimensional particle in a sech-squared potential well is a sech function.
Since the constant terms are are essentially aribtrary energy shifts, this means that, for some given interaction strength $g_{\mathrm{1d}}$ and particle number $N$, a trapping potential with the spatial dependency
\begin{equation}
U_S(x)=-\frac{1}{4\ell_{0}^2}\left(\frac{\hbar^{2}}{m} + g_{\mathrm{1d}}N\right)\mathop{\mbox{sech}^{2}}(x/2\ell_{0})
\end{equation}
will mean that the energy minimising stationary solution of 
\begin{equation}
-\frac{\hbar^{2}}{4m\ell_{0}^{2}}\psi=-\frac{\hbar^{2}}{2m}\frac{d^{2}\psi}{d x^{2}}+U_{S}\psi{+}g_{\mathrm{1d}}N|\psi|^{2}\psi, 
\end{equation}
is also given by Eq.~(\ref{Eq:SolitonWavefunction}). Hence, if our initial state is taken as the ground state of a repulsive interacting quasi one-dimensional condensate of $N$ particles, with interaction strength $g_{\mathrm{1d}}$ and trapping potential $U_{S}$, then instantaneously changing $g_{\mathrm{1d}}\rightarrow-g_{\mathrm{1d}}$ and $U_{S}\rightarrow 0$ should not produce any dynamics in the density profile of the wavefunction. It is the purpose of the remainder of the work to realistically formalize this scheme for future ultracold atom experiments.

\section{\label{sec:se}Soliton Engineering in 3D}
\subsection{Determination of the 3D initial trapping potential}
\noindent To account for realistic settings we need to construct a fully three-dimensional potential landscape as a test of the validity of the proposed method. The derivation of such a potential can be accomplished by considering, for a given target state, what potential would generate this state within the framework of the three-dimensional Gross--Pitaevskii equation. Again, realising that the ground state wavefunction in three dimensions is nodeless, such a potential can be written as
\begin{equation}\label{eqn:3dpot}
U\{\psi_{\rm T}({\bf r})\}=\mu+\frac{\hbar^2}{2m\psi_{\rm T}({\bf r})}\nabla^{2}\psi_{\rm T}({\bf r})-g|\psi_{\rm T}({\bf r})|^2.
\end{equation} 
Equation \eqref{eqn:3dpot} introduces the target state $\psi_{\rm T}({\bf r})$ and $g=4\pi\hbar^2|a_s|/m$ where again $m$ is the atomic mass, $a_s$ is the $s$-wave scattering length, and $\mu$ is the chemical potential of the atomic cloud. The choice of the target state is however not completely arbitrary. It must be a (locally) energy minimizing solution to the underlying Gross--Pitaevskii model (a consequence of attractive interactions is that there is in general no global energy minimum in the 3D Gross--Pitaevskii equation). As such, we use a variational wave function as the target state. Note that one could also use the full numerical solution to determine the target state. Using the variational solution proves to be advantageous, however, as the variational analysis provides an intuitive understanding of the underlying parameter space of the 3D problem. We consider the situation where the soliton is confined harmonically in both the radial (denoted $\rho^2=y^2+z^2$) and axial ($x$) directions. Then, a good soliton variational state is described by \cite{billam_2012}
\begin{equation}\label{eqn:tpsi}
\psi_{\rm S}({\bf r})=\frac{\sqrt{\gamma\kappa}k_{S}}{(4\pi\ell_S)^{1/2}}e^{-\kappa\gamma k_{S}^{2}\rho^{2}/2}\text{sech}(x/2\ell_S),
\end{equation}
\begin{figure}[t]
\includegraphics[width=\columnwidth]{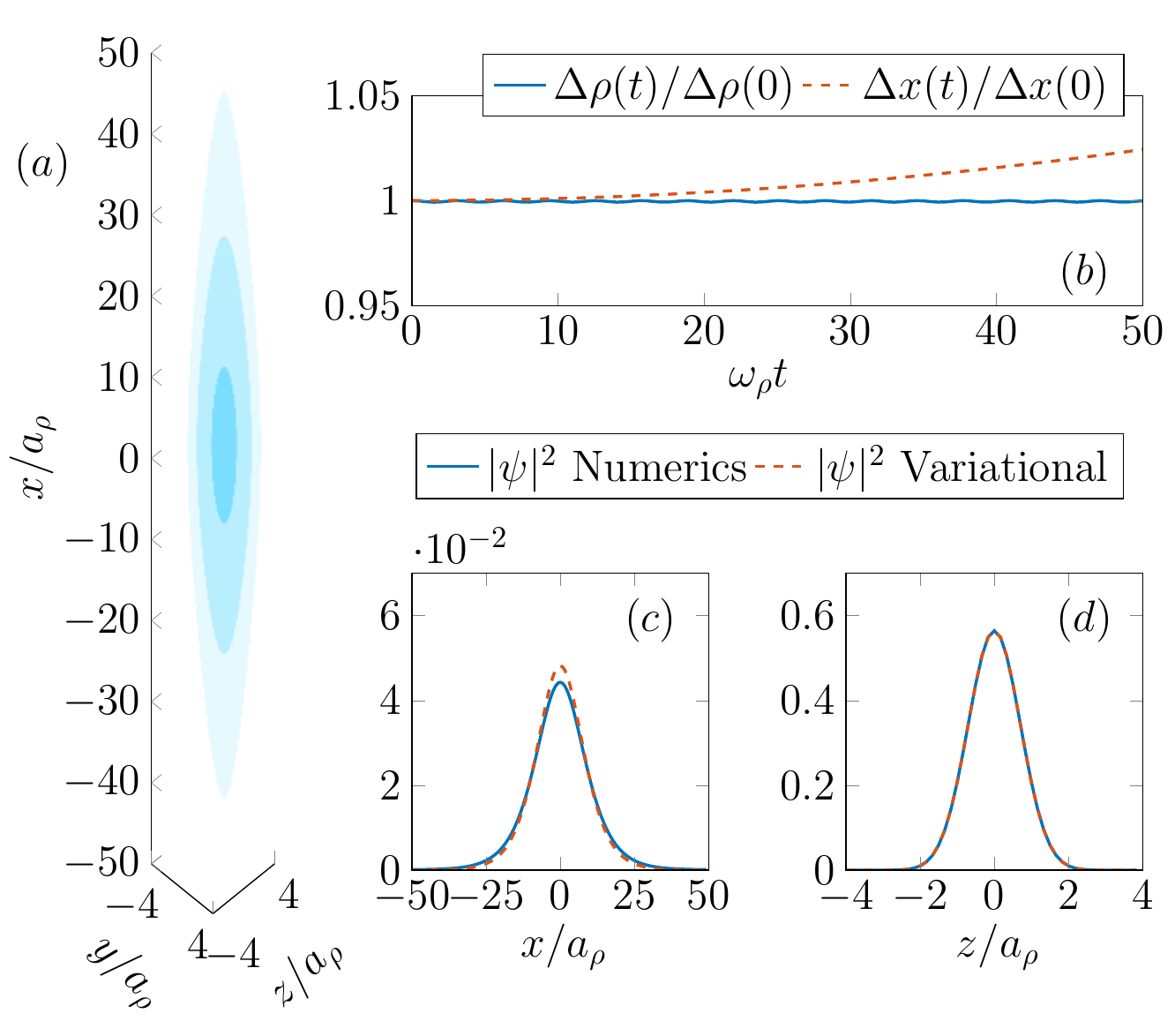}
\caption{\label{fig:3d}(Color online) Numerical simulation using the three dimensional tailored trapping potential. (a) shows isosurfaces of the bright soliton with density $|\psi({\bf r})|^2=10^{-4}, 10^{-3}, 10^{-2}$. Panel (b) shows the real-time standard deviations, $\Delta\rho(t)$ and $\Delta x(t)$, while (c) and (d) show comparisons between the projections of the three dimensional data and the variational wave function.}
\end{figure} 
where the parameters introduced by Eq.~\eqref{eqn:tpsi} are $\gamma=(a_{\rm\rho}^{2}/2a_sa_{x}N)^2$ which is a dimensionless ratio of the soliton and harmonic length scale, and $\kappa=\omega_{\rho}/\omega_{x}$ which defines the ratio of the radial and axial harmonic trapping frequencies, while $a_{x}$ and $a_\rho$ give the axial and radial harmonic length scales of the three dimensional trap. The remaining parameters are $\ell_{S}$ and $k_S$, which define the axial and radial variational length and inverse lengths respectively. Then, inserting the state given by Eq.~\eqref{eqn:tpsi} into the energy functional associated with Eq.~\eqref{eqn:3dpot} and minimizing the resulting energy with respect to the variational parameters leads to a cubic equation for $\ell_{S}$. The only physically relevant solution is given by
\begin{equation}\label{eqn:ls}
\frac{\ell_{S}}{\ell_0}{=}\frac{[\chi(u)]^{1/2}(k_S\ell_0)^{2/3}}{2^{11/6}(\pi\gamma)^{2/3}}\left(\left\{\left[\frac{2}{\chi(u)}\right]^{3/2}{-}1\right\}^{1/2}{-}1\right),
\end{equation}
where the soliton length scale $\ell_0$ is defined as in Section \ref{sec:pe}, and $u=\gamma(k_{S}\ell_{0})^{-4}$. Then, the function $\chi(u)=\chi_{+}(u)+\chi_{-}(u)$, where $\chi_{\pm}(u)$ is defined as
\begin{equation}\label{eqn:chi}
\chi_{\pm}(u)=\left[1\pm\left(1+\frac{1024\pi^2u^2}{27}\right)^{1/2}\right]^{1/3}.
\end{equation}
Meanwhile, the inverse radial variational length is obtained from
\begin{equation}\label{eqn:ks}
k_{S}=\frac{1}{\ell_0}\bigg(\frac{6\kappa\gamma\ell_S}{6\kappa\gamma\ell_S-\ell_0}\bigg)^{1/4}.
\end{equation}
Values for $k_{S}$ and $\ell_{S}$ are determined from Eq.~\eqref{eqn:ls} and Eq.~\eqref{eqn:ks} simultaneously, for which a straightforward numerical procedure exists \cite{billam_2012}.

Finally, inserting the target state given by Eq.~\eqref{eqn:tpsi} into Eq.~\eqref{eqn:3dpot} we arrive at the desired trapping potential 
\begin{align}\nonumber
U\{\psi_{\rm S}({\bf r})\}=&-\frac{\hbar^2}{ma_{\rho}^{2}}\frac{\ell_{0}^{2}}{4\gamma\kappa\ell_{S}^{2}}\left[1+2\ell_0\ell_S k_{S}^{2}e^{-k_{S}^{2}\rho^2}\right]\text{sech}^{2}\left(\frac{x}{2\sqrt{\gamma\kappa}\ell_S}\right)\\&+\frac{\hbar\omega_{\rho}}{2}\ell_{0}^{2}k_{S}^{4}\rho^2.
\label{eqn:sol3dpot}
\end{align}
To experimentally create the soliton given by Eq.~\eqref{eqn:tpsi}, one would first produce a condensate with the potential defined by Eq.~\eqref{eqn:sol3dpot} in the presence of repulsive interactions. Then, a quench is applied to the system by switching off the potential of Eq.~\eqref{eqn:sol3dpot}, changing the sign of the mean-field interactions from repulsive to attractive (maintaining the magnitude of $a_s$) and imposing the radial and axial trapping potential given by
\begin{equation}
U_{\rm trap}({\bf r})=\frac{1}{2}m\bigg(\omega_{x}x^2+\omega_{\rho}^{2}\rho^2\bigg).
\end{equation}
These precise conditions yield the required three-dimensional bright soliton state with minimal induced dynamics of the atomic cloud.

Figure \ref{fig:3d} shows an example of the ground state and real time dynamics obtained from having the initial trapping potential described by Eq.~\eqref{eqn:sol3dpot}. The numerical procedure used to procure this solution for a given scattering length and trapping geometry involves using an iterative biconjugate gradient scheme to find the ground state solution on the repulsive ($a_s>0$) side. Imaginary time propagation is insufficient here as it fails to eliminate low lying positive momentum modes that are present in the axial part of the three dimensional wave function. The real time dynamics are then handled using the XMDS2 software \cite{dennis_2013}. The example shown in Fig.~\ref{fig:3d} is for the choice of parameters $\kappa=400$ and $\gamma=0.1$, which corresponds to $k_{S}\ell_0\simeq0.65$ and $\ell_S/\ell_0\simeq1.001$. Panel (a) depicts isosurfaces of the three dimensional ground state density distribution, for $|\psi({\bf r})|^2=10^{-4}, 10^{-3}, 10^{-2}$; higher density regions have darker shading, and lower density ones have lighter shading. Then, panel (b) shows the standard deviation of the bright soliton in the axial $x$ and radial $\rho$ coordinate directions calculated during real-time evolution. The lower two panels, (c) and (d) depict comparisons of the one dimensional densities computed numerically from $n_{\mathrm{1d}}(x)=\int \mathop{dy}  \int \mathop{dz} |\psi(x,y,z)|^2$ to the analytical form of $\psi_{\rm S}({\bf r})$ [Eq.~\eqref{eqn:tpsi}]. The left panel (c) depicts the axial (soliton) density, while panel (d) depicts part of the radial form of the variational state.

The three-dimensional simulation of Eq.~\eqref{eqn:sol3dpot} demonstrates the insensitivity of the soliton radial dynamics, since $\Delta\rho(t)$ [solid blue, Fig.~\ref{fig:3d}(b)] is almost independent of time. Hence, in the sections that follow we work in a dimensionally reduced scenario to probe the axial dynamics of the bright soliton state.   

\subsection{\label{sec:q1d}Quasi-One-Dimensional Dynamics}
\noindent It is interesting to compare the variational 1D solution, where the radial dynamics of the cloud are effectively frozen out, such that any dynamics of the cloud are well described by an effective quasi-one-dimensional description, to the pedagogical model discussed in Sec.~\ref{sec:pe}. In this case, the counterpart of Eq.~\eqref{eqn:3dpot} is given by
\begin{equation}\label{eqn:1dpot}
U_{\rm 1D}\{\psi_{\rm T}(x)\}=\mu+\frac{\hbar^2}{2m\psi_{\rm T}(x)}\frac{d^2\psi_{\rm T}(x)}{d x^2}-g_{\mathrm{1d}}|\psi_{\rm T}(x)|^2,
\end{equation}
where $g_{\mathrm{1d}}=g/2\pi a_{\rho}^2$ is the scaled one-dimensional interaction parameter, and $\psi_{\rm T}(x)$ denotes the target state. We consider the situation where there is a harmonic trap, where the units of frequency are given by $\omega_{x}=\hbar/m\ell_{0}^{2}$, which accompanies the mean-field interactions. Then, an energy-minimizing state in this case can be written
\begin{equation}\label{eqn:psi1d}
\psi_{\rm S}(x)=\frac{1}{2\sqrt{\ell_S}}\text{sech}(x/2\ell_S)
\end{equation}
here the $\ell_S$ appearing in Eq.~\eqref{eqn:psi1d} is the one-dimensional variational length scale that can be computed analytically for a given scattering length from 
\begin{equation}\label{eqn:ls1d}
\frac{\ell_{S}}{\ell_0}=\frac{\sqrt{\chi(\gamma)}}{2^{11/6}(\pi\gamma)^{2/3}}\left(\left\{\left[\frac{2}{\chi(\gamma)}\right]^{3/2}-1\right\}^{1/2}-1\right),
\end{equation}
and $\chi(\gamma)=\chi_{+}(\gamma)+\chi_{-}(\gamma)$ is defined through Eq.~\eqref{eqn:chi}, as stated previously. In the limit $\gamma\rightarrow0$ Eq.~\eqref{eqn:ls1d} reduces to $\ell_S\rightarrow1$, recovering the integrable limit. Then, the quasi-one-dimensional initial trapping potential is computed by inserting Eq.~\eqref{eqn:psi1d} into Eq.~\eqref{eqn:1dpot} which, after dropping constant energy terms becomes
\begin{equation}\label{eqn:sol1dpot}
U_{\rm 1D}\{\psi_{\rm S}(x)\}=-\frac{\hbar^2}{2m a_{\rho}^{2}}\frac{N^2a_{s}^{2}}{a_{\rho}^{2}}\left(\frac{\ell_{0}^{2}}{\ell_{S}^{2}}+\frac{\ell_0}{\ell_S}\right)\text{sech}^{2}(x/2\ell_S).
\end{equation} 
To quantify how robust the proposed protocol is, we numerically solve for the ground state of Eq.~\eqref{eqn:sol1dpot} in the presence of repulsive mean-field interactions, then use this state as the initial condition for real-time propagation with attractive interactions and a harmonic trapping potential. A useful measure to investigate the fidelity of the states generated by Eq.~\eqref{eqn:sol1dpot} is the standard deviation $\Delta x(t)=\sqrt{\langle x^2(t)\rangle-\langle x(t)\rangle^{2}}$ which we would in general like to be independent of time.
\begin{figure}[t]
\includegraphics[width=\columnwidth]{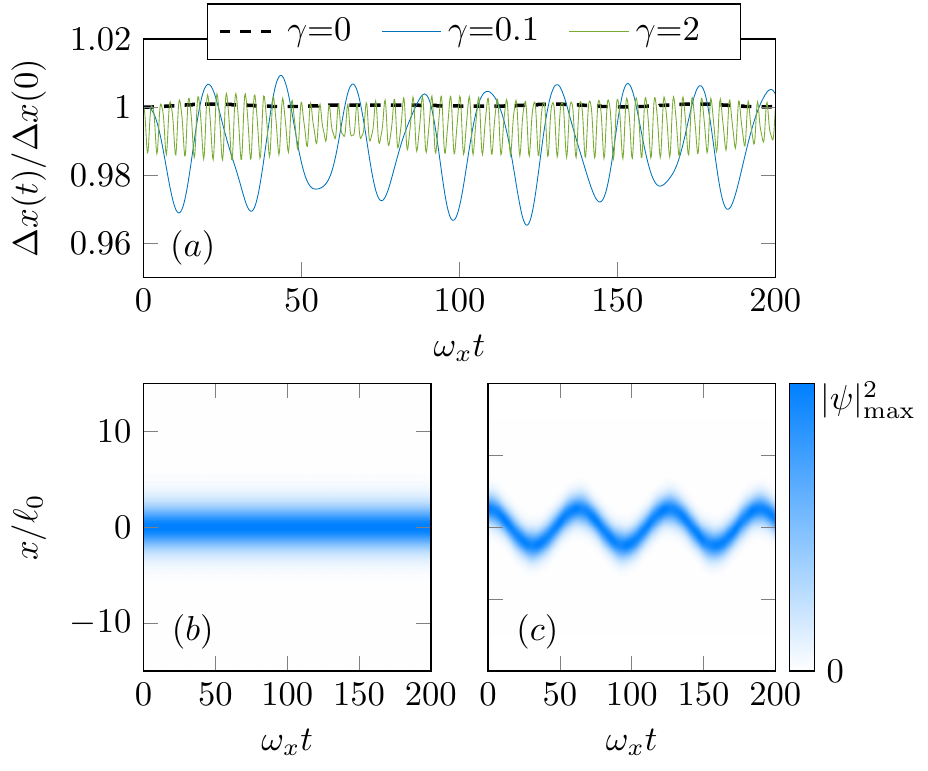}
\caption{\label{fig:1d}(Color online) Numerical solutions obtained from the time-dependent Gross--Pitaevskii equation assuming the quasi-one-dimensional initial trapping potential, Eq.~\eqref{eqn:sol1dpot}. Panel (a) shows the standard deviation of the state during real time dynamics for three different values of $\gamma=0, 0.1, 2$. Panels (b) and (c) show the density data $\psi(x,t)$ for $\gamma=0$, (b) and $\gamma=2$, (c). In (c), the minimum of the potential has been displaced by $5\ell_0/2$. The units of time are defined through the quantity $\omega_x=\hbar/m\ell_{0}^2$.}
\end{figure}
Figure \ref{fig:1d} shows example dynamics for three different values of $\gamma$. Each dataset has been scaled to its initial value for ease of comparison. For $\gamma=0$ (black dashed line) the width of the state is independent of time, indicating that the protocol has generated the desired target state. For $\gamma>0$ the width undergoes small amplitude oscillations, which can be attributed to the underlying approximate nature of the variational state. We observe that the amplitude of these oscillations decreases with increasing $\gamma$, but the effective oscillation frequency increases, when comparing $\gamma=0.1$ (blue solid line) with $\gamma=2$ (green solid line). The two lower panels in Fig.~\ref{fig:1d} show example dynamics in the form of $|\psi(x,t)|^2$. Fig.~\ref{fig:1d}(b) corresponds to $\gamma=0$, while Fig.~\ref{fig:1d}(c) corresponds to $\gamma=2$. In this second example, the minimum of the initial trapping potential was offset by $5\ell_0/2$, demonstrating the stability of this state's center-of-mass motion in the harmonic trapping potential.

\subsection{\label{sec:fluc}Sensitivity to fluctuations}
\noindent So far, we have considered a protocol starting from the (numerically obtained)
ground state of the tailored initial trapping potential. While it is feasible for experiments to
reach low temperatures with very low thermal fraction, it is unavoidable that
the initial state contains some fluctuations; although as $T \rightarrow 0$
thermal fluctuations diminish, quantum fluctuations remain unavoidable. It is
therefore useful to characterize the sensitivity of the protocol to these
fluctuations. Working within the quasi-one-dimensional description, we adopt
the truncated Wigner (TW) method \cite{steel_1998, sinatra_2002, blakie_2008}
to characterize this.  The TW method has previously been applied to bright
solitons, for example in Refs.~\cite{dabrowska_wuster_2009, martin_2012,
haine_2018}, and we follow the approach of Ref.~\cite{martin_2012} to simulate
fluctuations at $T=0$. First we numerically solve for the ground state in the initial trapping
potential as defined in Eq.~(\ref{eqn:sol1dpot}), and then we solve numerically for Bogoliubov--de Gennes modes
$u_j(x)$, $v_j(x)$ orthogonal to this ground state \cite{blakie_2008}, assuming periodic boundary conditions over length $25.6 \ell_0$. For our
stochastic initial conditions we seed $M=256$ modes with on average half a
particle of noise sampled according to the prescription 
\begin{equation}
\psi_i(x,0) = \psi_\mathrm{GP}(x,0) + \sum_j^M \left[\beta_{i,j} u_j(x) + \beta_{i,j}^* v_j^*(x) \right],
\end{equation}
where the Gaussian complex random variables $\beta_{i,j}$ obey
\begin{align}
\langle \beta_{i,j} \rangle = \langle \beta_{i,j} \beta_{i,k} \rangle &= 0,\\
\langle \beta_{i,j}^* \beta_{i,k} \rangle &= \frac{1}{2} \delta_{jk}.
\end{align}

We evolve $N_t = 200$ stochastic initial conditions forward in time using the
quasi-one-dimensional Gross--Pitaevskii equation, having instantaneously removed
the initial trapping potential and switched the sign of the interaction at time $t=0$
according to the protocol. These stochastic trajectory simulations are
performed in XMDS2 \cite{dennis_2013} using a Fourier basis. In each
trajectory, indexed by $i$, we track the location of the soliton center of mass
as $\langle x(t) \rangle_i$ \cite{martin_2012}. Note that here $\langle \cdot \rangle_i$ denotes an average over a single trajectory as in Sec.~\ref{sec:q1d}. In comparison to the
pure Gross--Pitaevskii evolution described above, we find that the
per-trajectory standard deviation $\Delta x_i (t) = \sqrt{\langle x^2(t)\rangle_i - \langle x(t) \rangle_i}$ is not a particularly useful measure to
characterize the fidelity of the evolution for the high density regions of
interest close to the trap center; high values of the variance can be generated
by very small density values far from the trap center. Instead, we
least-squares fit each trajectory with a function of the form
\begin{equation}\label{eqn:fit}
\rho_i(x,t) = A_i(t) \,\mathrm{sech}^2 \left(\frac{x-\langle x(t)\rangle_i}{2 \ell_{S,\,i}(t)} \right),
\end{equation}
in order to extract the effective width $\ell_{S,\,i}(t)$. The quantity $A_i(t)$ appearing in Eq.~\eqref{eqn:fit} is a fitted, time-dependent amplitude. Operating under the
interpretation that one TW trajectory is in a sense comparable to one
experimental realization \cite{blakie_2008}, this fitting procedure is in fact
rather similar to likely procedures for extracting information about the
dynamics from experimental measurements.
\begin{figure}[t]
\includegraphics[width=\columnwidth]{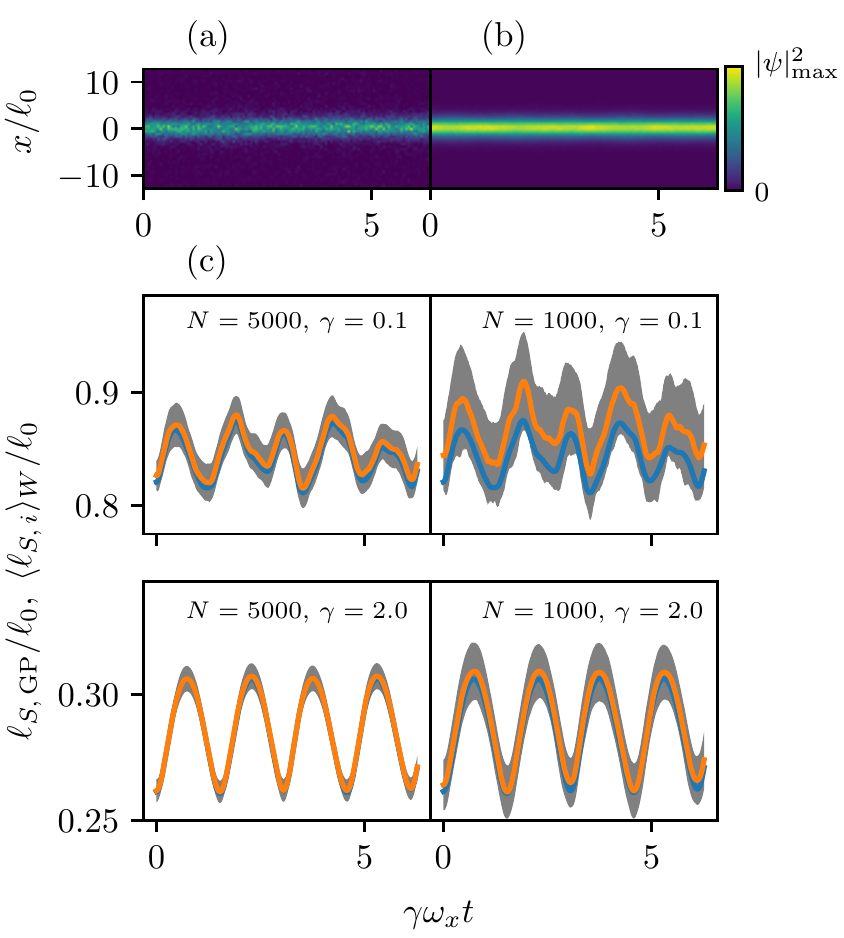}
\caption{\label{fig:tw}(Color online) Sensitivity of the quasi-one-dimensional model to
fluctuations in a truncated Wigner treatment at $T=0$.  We show in (a) the
density of a typical trajectory $|\psi_i|^2$, and in (b) the average density
over trajectories $\langle |\psi_i|^2 \rangle_W$, for the case $\gamma=0.1$,
$N=1000$.  In (c) we show the average effective widths of the soliton, obtained
by fitting a sech$^2$ profile to each trajectory (see text).  Specifically, we
compare the average over truncated Wigner trajectories ($\langle \ell_{S,\,i}
\rangle_W$, orange, grey shaded area denotes $\pm 1$ standard deviation) to the
width obtained by fitting the pure Gross--Pitaevskii solution
($\ell_\mathrm{GP}$, blue). Parameters are indicated in each subplot.}
\end{figure}
We find that variations in the soliton center of mass between trajectories are
very small compared to the width of the soliton $\ell_0$ ($\lesssim 5\%$).
Figure \ref{fig:tw} summarizes the results. Importantly, in
Fig.~\ref{fig:tw}(c) we show the width obtained by sech$^2$ profile fitting for
both weak ($\gamma = 0.1$) and strong ($\gamma = 2.0$) axial confining
potentials and for ground state atom numbers of $N=1000$ and $N = 5000$. In all
cases the width obtained by sech$^2$ profile fitting when averaged over all
trajectories, $\langle \ell_{S,\,i} \rangle_W$, is close to the pure
Gross--Pitaevskii result, although deviation from it it is more noticeable for
$N=1000$.  Overall, the results show that initial fluctuations for cold initial
conditions will not qualitatively affect the proposed protocol. Furthermore,
they show that fitted soliton widths extracted from noisy data (theoretical or
experimental) are in good quantitative agreement with the pure Gross--Pitaevskii
result.
\begin{figure}[t]
\includegraphics[width=\columnwidth]{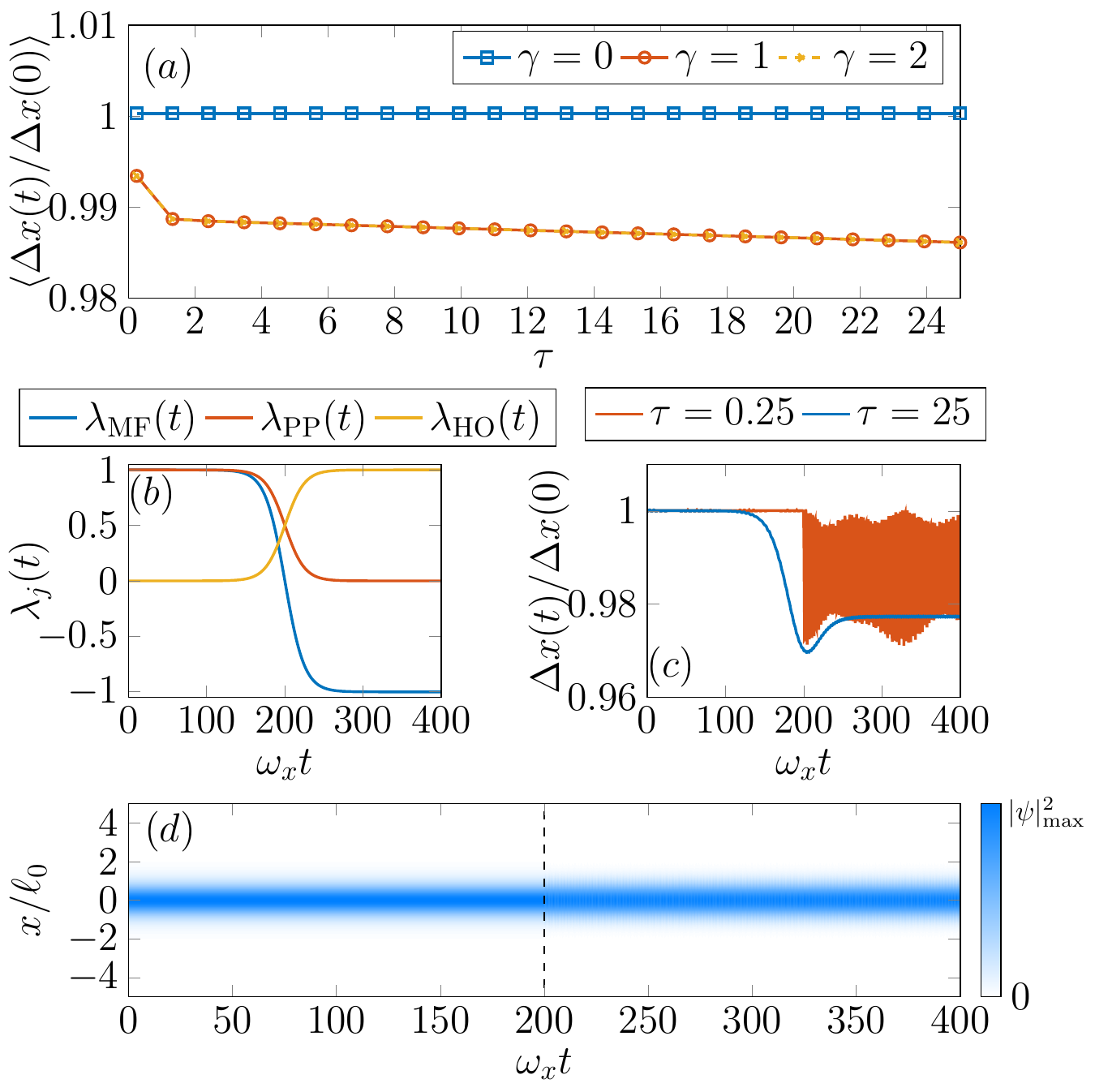}
\caption{\label{fig:quench}(Color online) Finite-time switching of the initial trapping potential. Panel (a) shows the fidelity of the switch as a function of the switch time scale $\tau$. The scaling functions given by Eqs.~\eqref{eqn:lmf}--\eqref{eqn:lpp} are plotted for a slow (adiabatic) switch from repulsive to attractive interactions. The sizes of the wave packet $\Delta(x)$ are shown in panel (c), for a fast (orange curve) and a slow (blue curve) switch. The final panel (d) displays a space-time plot for a soliton experiencing a fast switch, $\tau=0.25$. The dashed line indicates the center of the switch, $t=t_0$.}
\end{figure}
\subsection{\label{sec:qd}Quench Dynamics}
\noindent We can interrogate the robustness of the method further by quantifying how switching the various interaction terms in finite time affects the fidelity of the final state. To do this, we perform a time-dependent scaling of the mean-field scattering length, harmonic trap frequency and initial trapping potential amplitude by
\begin{subequations}
\begin{align}
a_{s}&\rightarrow\lambda_{\rm MF}(t)a_{s},\\
\omega_{\rm ho}&\rightarrow\lambda_{\rm HO}(t)\omega_{\rm ho},\\
U_{\rm 1D}&\rightarrow\lambda_{\rm PP}(t)U_{\rm 1D},
\end{align}
\end{subequations}
where the time-dependent scaling parameters $\lambda_{j}(t)$ are defined as
\begin{subequations}
\begin{align}\label{eqn:lmf}
\lambda_{\rm MF}(t)&=-\text{tanh}([t-t_0]/\tau),\\
\lambda_{\rm HO}(t)&=\frac{1}{2}\bigg\{\text{tanh}([t-t_0]/\tau)+1\bigg\},\\
\lambda_{\rm PP}(t)&=-\frac{1}{2}\bigg\{\text{tanh}([t-t_0]/\tau)-1\bigg\},\label{eqn:lpp}
\end{align}
\end{subequations}
here the time scale of the switch is denoted $\tau$ while $t_0$ is the effective ``center'' of the finite time switch. This simple protocol allows us to investigate the effect of going from a fast ($\tau\ll1$) to a slow ($\tau\gg1$) switching time scale. One would naively expect a rapid, violent change in the physical parameters to cause violent dynamics, while a slow quasi-adiabatic switch should allow the wave function time to evolve into a stable end state. We can quantify the switching from slow to fast by defining the fidelity in terms of the time-averaged width of the wavepacket, which is defined as
\begin{equation}\label{eqn:fid}
\langle\Delta x(T)\rangle=\frac{1}{T}\int_{0}^{T}dt\ \Delta x(t).
\end{equation}
Figure \ref{fig:quench} demonstrates the switching protocol for the quasi one-dimensional model, encapsulated by Eq.~\eqref{eqn:sol1dpot}. The top panel (a) displays the fidelity [Eq.~\eqref{eqn:fid}] for three different values of $\gamma$. In the pedagogical limit $\gamma=0$ one recovers a perfect fidelity, which is connected to the underlying fact that the integrability of the system has been restored. For more realistic scenarios where $\gamma>0$, the fidelity although no longer perfect is still very good, with only a ${\sim}1\%$ difference from the $\gamma=0$ case. The two middle panels, (b) and (c) of Fig.~\ref{fig:quench} show a specific example of the protocol. In (b), the three finite time scaling relations, Eqs.~\eqref{eqn:lmf}--\eqref{eqn:lpp} are plotted for $\tau=25$. The final panel Fig.~\ref{fig:quench}(d) shows example dynamics for a fast switch $(\tau=0.25\text{ and }\gamma=1)$. Here the rapid oscillations of the amplitude of the soliton can just be seen after the quench, indicated by the dashed line.
   
\section{\label{sec:con}Conclusion}
\noindent In this work we have proposed a novel method for engineering solitons with good fidelity of the final bright soliton state. This is based on using appropriately tailored trapping potentials for the initially repulsively interacting atoms to condense into, for example by suitable generalization of laser generated painted potentials, the ground state of which yields the desired target soliton mode. We showed that in three dimensions this scheme can be used to generate a quasi one-dimensional soliton-like state with good fidelity. We have investigated the role of fluctuations via the Truncated Wigner approach, where we demonstrated that the dynamics of the cloud are not dramatically affected by the presence of noise. We also investigated how the rate at which the interactions are changed from repulsive to attractive affects the target state. We found the fidelity of the final state to be almost independent of the time scale of the switching, however we observed that rapid low amplitude oscillations of the width of the wave packet occur when the switching is most rapid, which could cause atomic losses in a real experiment. Hence, a slower, more adiabatic switch would be favorable to stable soliton dynamics.   

This scheme is not limited to the particular example presented in this work. Indeed, in the future it would be interesting to investigate the generation of other ultracold atomic structures such as vortices and vortex lattices with this protocol, and also to generalise the methodology to systems comprising several components, such as higher spin models.     

\acknowledgements
\noindent We acknowledge useful discussions with Oliver J. Wales and Ana Rakonjac. MJE acknowledges support as an Overseas researcher under Postdoctoral Fellowship of Japan Society for the Promotion of Science. TPB acknowledges support from UK EPSRC grant EP/R021074/1. SAG acknowledges support from UK EPSRC grants EP/L010844/1 and EP/R002061/1. This work was supported by the Okinawa Institute of Science and Technology Graduate University. Data supporting this publication is openly available under an Open Data Commons Open Database License \cite{data}.

\end{document}